\documentclass[a4paper,11pt]{article}
\usepackage{pos}
\usepackage{subcaption}

\title{Scaling region of the 3D Ising universality class in finite temperature QCD}

\author[a]{Michele Caselle}
\author*[b]{Marianna Sorba}

\affiliation[a]{Department of Physics, University of Turin and INFN Turin\\
   Via Pietro Giuria 1, I-10125 Turin, Italy}

\affiliation[b]{SISSA and INFN Trieste\\
Via Bonomea 265, 34136 Trieste, Italy}

\emailAdd{caselle@to.infn.it}
\emailAdd{msorba@sissa.it}

\abstract{We introduce a universal combination of susceptibility and correlation length in the 3D Ising model, depending both on temperature and external magnetic field. Starting from a parametric representation of the equation of state, we study its behaviour close to the critical point. The results we derive can be used as a sort of “reference frame” to chart the scaling region of the 3D Ising universality class. More specifically, we focus on instances of Ising behaviour in finite temperature QCD and, among these, we are particularly interested on the critical ending point in the finite density, finite temperature QCD phase diagram. In this context, Monte Carlo simulations are not possible and it is particularly difficult to disentangle “magnetic-like” from “thermal-like” observables, thus an explicit charting of the critical region could be useful for a direct comparison of experimental results with QFT/Statmech predictions.}

\FullConference{%
 The 38th International Symposium on Lattice Field Theory, LATTICE2021
  26th-30th July, 2021
  Zoom/Gather@Massachusetts Institute of Technology
}

\begin{document}
\maketitle

\section{Introduction}

In this proceeding we present our latest work \cite{Caselle:2020tjz} aimed at improving the way to chart the scaling region of the Ising universality class, with particular attention to the three-dimensional case.\\
The Ising universality class plays a key role in theoretical physics due to its many realizations in different physical contexts, from condensed matter to high energy physics. From a statistical mechanics point of view, the Ising model describes the simplest critical system characterized by short-range interactions and undergoing a symmetry breaking phase transition, with a scalar order parameter. In the quantum field theory language, it is the simplest example of a unitary conformal field theory perturbed by only two relevant operators.\\
From conformal perturbation and bootstrap \cite{Kos:2016ysd}, several results are known on the behaviour of the model in three dimensions at the critical point or when a single perturbing operator is present. Nevertheless, the typical interesting regime in experiments is when both relevant perturbations are switched on, but far fewer predictions are available in this circumstance. The idea is therefore to study a suitable universal function which could give a description of the whole scaling region close to the Ising critical point.\\
Based on the universality hypothesis, the results we will derive apply not only for the standard three-dimensional Ising model, but for any physical realization of its universality class. We especially address examples of Ising behaviour in finite temperature QCD: in this framework, it is not easy to identify the relevant directions in the phase diagram corresponding to a ``thermal'' and a ``magnetic'' perturbation of the critical point and, in addition, numerical Monte Carlo simulations in a finite density regime suffer from the well-known sign problem, thus the introduction of a ``reference frame'' for the critical region could be a guideline in the interpretation of experimental results.

\section{3D Ising model and observable $\Omega$}

The spin model we are going to study is the Ising model on a cubic lattice, defined by the following energy function
\begin{equation}
    E(\{\sigma_i\})=-J\sum_{\langle ij\rangle} \sigma_i \sigma_j -H\sum_{i=1}^{N} \sigma_i\,,
    \label{Ising_energy}
\end{equation}
with spins $\sigma_i=\pm 1$ placed on the sites of the lattice, $J$ represents the coupling strength between them (we assume $J>0$) and $H$ is the externally applied magnetic field. For $H=0$ the model has a global $\textbf{Z}_2$ symmetry, which is spontaneously broken in a low temperature phase ($T<T_c$) characterized by the emergence of a non-zero spontaneous magnetization; conversely, in a high temperature phase ($T>T_c$) the $\textbf{Z}_2$ symmetry is restored. The two phases are separated by a critical point located at $H=0$ and $T=T_c$, that is actually the ending point of a line of first order phase transitions for $H=0$ and $T<T_c$. In the scaling region close to this critical point, a continuum limit of the spin model gives a quantum field theory described by the action
\begin{equation}
     S = S_{CFT}+t\int d^3 x \,\epsilon(x) +H\int d^3 x \,\sigma(x)\,.
     \label{Ising_QFT}
\end{equation}
Here $\epsilon(x)$ and $\sigma(x)$ are, respectively, the energy density ($\textbf{Z}_2$ even) and the spin ($\textbf{Z}_2$ odd) relevant operators which perturb the critical point. Indeed, deviations from the critical point are measured in terms of the reduced temperature $t=(T-T_c)/T_c$ and the magnetic field $H$, which are conjugated variables to the operators $\epsilon$ and $\sigma$ respectively. The term $S_{CFT}$ is the conformal-invariant action at the critical point.\\
From the order parameter of the model $\langle \sigma\rangle=M$ (i.e. the magnetization), we obtain the magnetic susceptibility $\chi=\partial M/\partial H$ and the exponential correlation length $\xi$, defined by the decay of the connected spin-spin correlator $\langle \sigma(x) \sigma(0)\rangle_c=\langle \sigma(x) \sigma(0)\rangle-\langle \sigma(0)\rangle^2 \sim e^{-|x|/\xi}$ as $|x|\to \infty$. These two observables are easily computable in numerical simulations and measurable in experiments, thus being the perfect ingredients to construct a universal combination of thermodynamic quantities that could describe the whole Ising scaling region, when both the $t$ and $H$ perturbations are present.\\
We shall hence study the following ratio
\begin{equation}
    \Omega=\left(\frac{\chi(t,H)}{\Gamma_-}\right) \left(\frac{\xi_-}{\xi(t,H)}\right)^{\gamma / \nu}\,,
    \label{Omega}
\end{equation}
where the normalization factors $\Gamma_-$ and $\xi_-$ denote the amplitudes of $\chi$ and $\xi$ along the first order phase transitions axis $t<0$, $H=0$. They are defined by the scaling behaviours $\chi \sim \Gamma_- (-t)^{-\gamma}$ and $\xi\sim \xi_- (-t)^{-\nu}$, with critical exponents $\gamma=(3-2x_{\sigma})/(3-x_{\epsilon})$ and $\nu=1/(3-x_{\epsilon})$ expressed in terms of the $\sigma$ and $\epsilon$ scaling dimensions, recently computed for the Ising model in three dimensions via a bootstrap approach \cite{Kos:2016ysd} as $x_{\sigma}=0.5181489(10)$ and $x_{\epsilon}=1.412625(10)$. This choice of normalization for the universal ratio $\Omega$ is again dictated by the fact that the coexistence line $t<0$, $H=0$ is usually simple to identify both numerically and in experiments.\\
Since we are interested in the description of the entire scaling region, it is useful to combine the two relevant perturbations, i.e. temperature and magnetic field, in the form of a scaling variable defined as
\begin{equation}
\eta=\frac{t}{H^{1/\beta\delta}}\,,
\label{eta}
\end{equation}
with $\beta=x_{\sigma}/(3-x_{\epsilon})$ and $\delta=(3-x_{\sigma})/x_{\sigma}$. Specifically, the limits in which only one perturbation is present ($t<0$, $H= 0$), ($t= 0$, $H\neq 0$), and ($t>0$, $H= 0$) correspond respectively to $\eta=-\infty$, $\eta=0$, and $\eta=+\infty$. In these three limits the function $\Omega$ can be written in terms of the standard universal amplitude ratios obtained with remarkable precision in Monte Carlo simulations \cite{Caselle:1997hs, Hasenbusch:2010ua} $Q_2=1.207(2)$, $\Gamma_+/\Gamma_-=4.714(4)$, and $\xi_+/\xi_-=1.896(3)$ as
\begin{equation}
    \begin{aligned}
      \Omega(\eta)&=  1 &\quad & \eta=-\infty\,,  \\
      \Omega(\eta)&= \frac{1}{Q_2} \left(\frac{\Gamma_+}{\Gamma_-}\right) \left(\frac{\xi_-}{\xi_+}\right)^{\gamma / \nu}   &\quad & \eta=0\,,  \\
      \Omega(\eta)&= \left(\frac{\Gamma_+}{\Gamma_-}\right) \left(\frac{\xi_-}{\xi_+}\right)^{\gamma / \nu} &\quad & \eta=+\infty\,.  \\
         \label{Omega_limits}
    \end{aligned}
\end{equation}
We then have numerically $\Omega(-\infty)=1$, $\Omega(0)=1.25(11)$, and $\Omega(+\infty)=1.355(71)$. These three values will be used later as benchmarks to test the reliability of our estimates for $\Omega$ at varying $\eta$.

\section{Parametric representation}

We address the problem making use of a parametric representation of the Ising model, as first introduced in \cite{Schofield:1969zza}. The critical equation of state is expressed in terms of two positive parameters $(R,\theta)$ according to
\begin{equation}
    \left\{
    \begin{aligned}
    M &= m_0 R^{\beta} \theta\,, \\
    t &= R (1-\theta^2)\,, \\
    H &= h_0 R^{\beta \delta} h(\theta)\,,
    \end{aligned}
    \right.
    \label{parametric_eos}
\end{equation}
where $m_0$ and $h_0$ are non-universal constants. The smallest positive zero of the unknown function $h(\theta)$ is called $\theta_0>1$, thus the lines $\theta=\theta_0$,  $\theta=1$, and $\theta=0$ are respectively identified with the axes ($H=0$, $t<0$), ($H\neq 0$, $t=0$), and ($H=0$, $t>0$). The physically interesting domain in the $(R,\theta)$ plane is then $0\le \theta \le \theta_0$ for every $R\ge 0$. Within this range, the function $h(\theta)$ must be analytic and odd in $\theta$ (because of the $\textbf{Z}_2$ symmetry of the model), therefore we can write it in general as a polynomial $h(\theta)=\theta+\sum_{n=1} h_{2n+1} \theta^{2n+1}$. The coefficients $h_{2n+1}$ are fixed by means of a variational approach \cite{Campostrini:2002cf}, and the polynomial is truncated up to the order that best fits the known values for the universal amplitude ratios. Following \cite{Tarko:1975zz}, a similar parametric representation can be also introduced for the correlation length, actually for its squared inverse which corresponds to the square mass of the underlying Ising quantum field theory. We use
\begin{equation}
    \xi^{-2}= R^{2\nu}a_0(1+c\theta^2)\,,
    \label{parametric_correlation_length}
\end{equation}
where $a_0$ is again a non-universal constant, while $c$ is extracted from the universal amplitude ratio $\xi_+/\xi_-$. Finally, the parametric expression for our function $\Omega$ turns out to be
\begin{equation}
    \Omega(\theta)= \Omega_0 \frac{(1-\theta^2+2\beta\theta^2)(1+c\theta^2)^{\frac{\gamma}{2\nu}}}{2\beta\delta\theta h(\theta)+(1-\theta^2)h'(\theta)}\,,
    \label{parametric_Omega}
\end{equation}
with normalization
\begin{equation}
    \Omega_0= \frac{(1-\theta_0^2)h'(\theta_0)}
{(1-\theta_0^2+2\beta\theta_0^2)(1+c\theta_0^2)^{\frac{\gamma}{2\nu}}}\,.
    \label{parametric_Omega_0}
\end{equation}
It is worth stressing that $\Omega$ is a monotonic decreasing function of $\theta$, hence invertible: from a given experimental estimate of $\Omega$, a precise value of the scaling variable $\theta$ can be derived. In this sense the universal ratio $\Omega$ could help to chart the critical region of the three-dimensional Ising universality class. Generally, in practical applications the function $\Omega$ is more conveniently expressed as a function of $\eta$ (see Equation \ref{eta}). To this purpose, we can write expansions of the scaling variable $\theta$ as a function of $\eta$ around the three singular points $\eta=0,\pm \infty$ and then get three corresponding series of $\Omega$ expressed in terms of $\eta$. See \cite{Caselle:2020tjz} for more details on this change of variables.

\section{Results}

\begin{figure}[t]
    \centering
    \begin{subfigure}[h]{0.47\textwidth}
        \includegraphics[width=\textwidth]{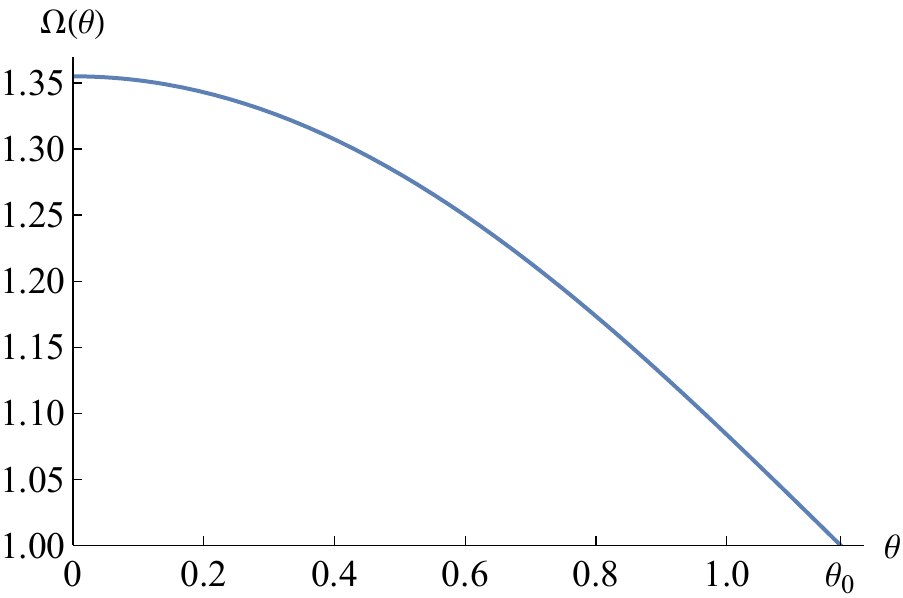}
    \end{subfigure}\hfill
    \begin{subfigure}[h]{0.47\textwidth}
        \includegraphics[width=\textwidth]{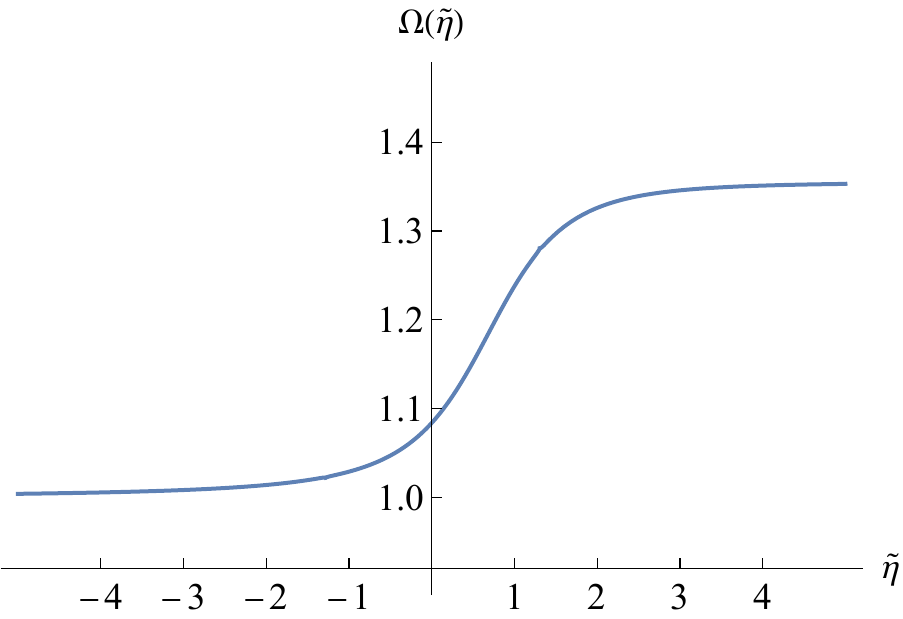}
    \end{subfigure}
    \caption{\textit{Left panel:} Result for $\Omega(\theta)$ according to Equation \ref{parametric_Omega} in the physical domain $0\le \theta\le \theta_0$.\\
    \textit{Right panel:} Result for $\Omega(\tilde{\eta})$ with $\tilde{\eta}= (h_0)^{1/\beta\delta}\, \eta$, in the three limits $\tilde{\eta}=0, \pm \infty$ where the scaling variable $\theta$ can be expanded in power series as a function of $\eta$.}
    \label{Figure1}
\end{figure}

Reliable estimates of the coefficients in the polynomial approximation of $h(\theta)$ are obtained in \cite{Campostrini:2002cf} up to the $7$-th order as
\begin{align}
    h(\theta)&= \theta+ h_3\, \theta^3+ h_5\, \theta^5 +h_7\, \theta^7 + O(\theta^9)\nonumber\\
    &= \theta-  0.736743 \, \theta^3+ 0.008904 \, \theta^5 - 0.000472\, \theta^7+ O(\theta^9)\,,
    \label{polynomial}
\end{align}
and the smallest positive root is $\theta_0^2=1.37861$. Using the Monte Carlo value of $\xi_+/\xi_-$ \cite{Hasenbusch:2010ua} we obtain $c=0.0416$ to be inserted in the parametric formula for the correlation length. Finally we test the consistency of our parametric representation of $\Omega$ by computing the universal amplitude ratios $\Gamma_+/\Gamma_-$, $Q_2$ and comparing them with the numerical estimates detailed above. Since the agreement is good (with a difference of the order of $1$-$3\%$), we can trust the result for $\Omega(\theta)$ and, correspondingly, that for $\Omega(\eta)$, both depicted in Figure \ref{Figure1}. More specifically, we notice that $\Omega$ in the three singular limits $\eta=0, \pm \infty$ (or, equivalently, at $\theta=0, 1, \theta_0$) matches the benchmark values described in the previous section.

\section{Applications in finite temperature QCD}

\begin{figure}[t]
    \centering
    \begin{subfigure}[h]{0.38\textwidth}
        \includegraphics[width=\textwidth]{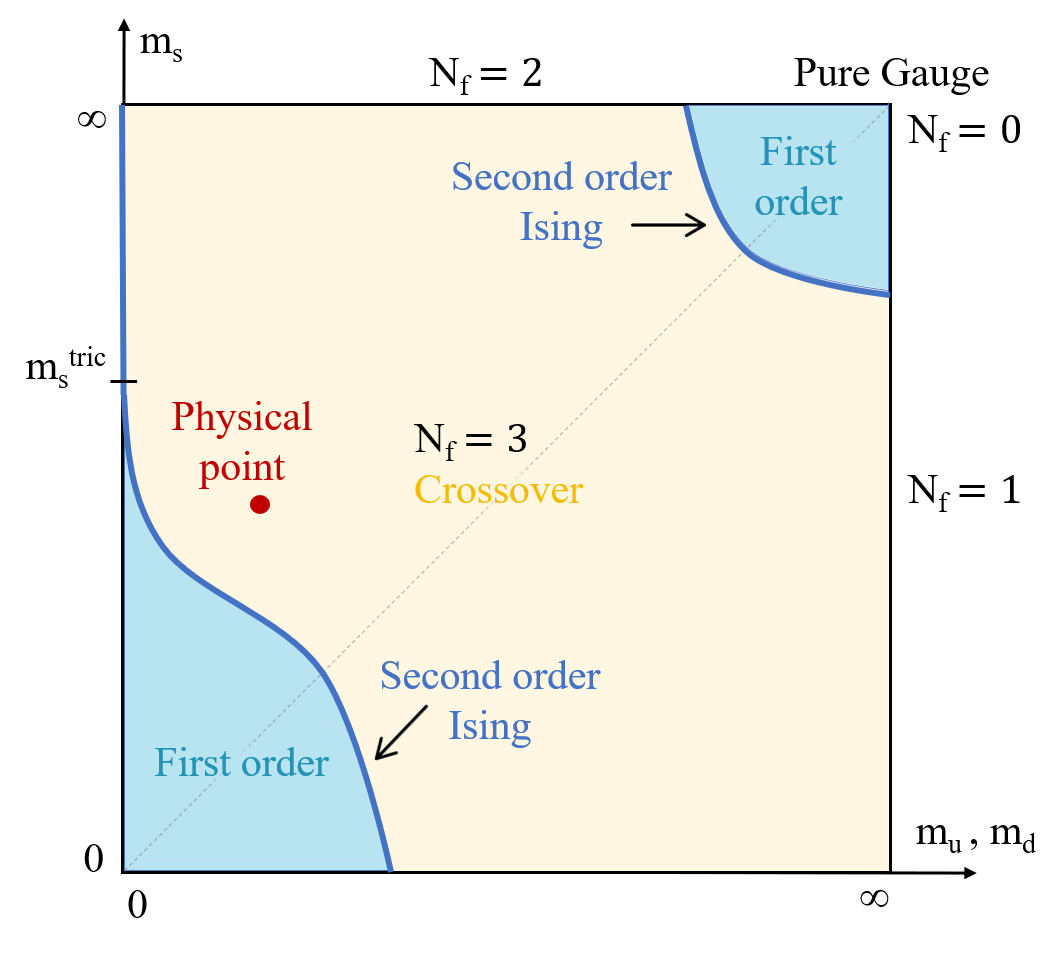}
    \end{subfigure}\hfill
    \begin{subfigure}[h]{0.58\textwidth}
        \includegraphics[width=\textwidth]{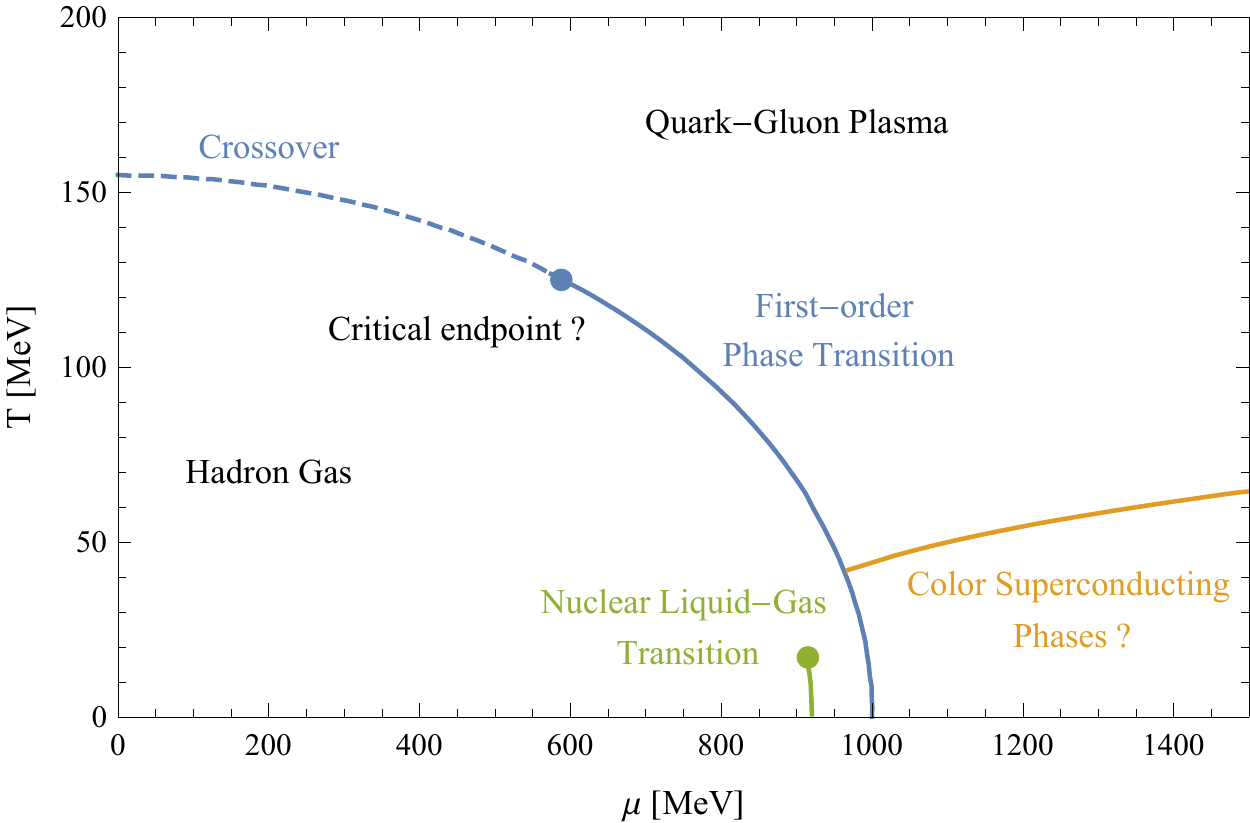}
    \end{subfigure}
    \caption{\textit{Left panel:} Columbia plot. We focus on the two second order phase transitions lines, belonging to the 3D Ising universality class. The top-right critical line is at the end of a deconfinement transition region, while the bottom-left one delimits a chiral transition region. The physical point lies in the crossover region between them.\\
    \textit{Right panel:} QCD phase diagram at finite baryon density. The precise location of the critical ending point is still uncertain, but it is believed to belong to the 3D Ising universality class.}
    \label{Figure2}
\end{figure}

In the phase diagram of finite temperature QCD one finds a rich pattern of phase transitions as the values of the quarks masses are varied. The situation is well summarized by the so-called Columbia plot (see the left panel in Figure \ref{Figure2}). For intermediate values of the masses of the three quark flavours, there is no phase transition but only a smooth crossover between the quark-gluon plasma phase at high temperature and the confined phase at low temperature. However, in the limit of infinite masses (top-right corner of the plot) the model becomes a pure $SU(3)$ gauge model, exhibiting a first order deconfinement phase transition. Decreasing the masses of the quarks, this first order region ends into a critical line of second order phase transitions which are expected to be of the three-dimensional Ising type. Similarly, in the limit of vanishing quarks masses (bottom-left corner of the plot) there is a region of first order phase transitions associated to chiral symmetry restoration which also ends into a critical line of second order Ising phase transitions.\\
Since the physical point, which lies in the crossover region, is not too far from this second Ising line, it is well possible that our results could be used to describe the finite temperature crossover between the confining and the plasma phases in QCD, even with the physical values of the quarks masses.\\
Finally, our results could be helpful also to describe the QCD critical point in the finite temperature, finite density phase diagram. This regime is relevant for the description of heavy-ion collision experiments \cite{Lacey:2006bc}, and cannot be studied by Monte Carlo simulations due to the well-known sign problem. It is nowadays generally believed \cite{Stephanov:2006zvm, parotto2020qcd} that a first order phase transitions line ending into an Ising critical point separates the quark-gluon plasma phase (at high $T$ and high $\mu$) from the hadronic phase (at low $T$ and low $\mu$), as illustrated in the right panel of Figure \ref{Figure2}. Even if the location of this critical ending point is still under discussion, the fact that it should belong to the three-dimensional Ising universality class makes our charting of this scaling region particularly useful in the search of its position. 

\section{Conclusions}

We showed how a suitable, and easily accessible in simulations and experiments, combination of thermodynamic quantities may give a precise charting of the Ising scaling region, in presence of both relevant perturbations. The universal predictions we derived are important to test numerical and experimental observations for any possible realization of the three-dimensional Ising universality class. In particular, we detailed the application of our results in finite temperature QCD, where an increasing amount of results coming from recent collision experiments needs to be compared with theoretical predictions, in order to get evidence on the conjectured location of the critical ending point in the $(T,\mu)$ phase diagram. The scaling region charting performed here, as well as other new complementary results \cite{Kadam:2020utt, Pan:2021svt, Pan:2021fus, Kotov:2021hri, Kotov:2021rah}, moves in this direction.\\
The goal of constructing a ``reference frame'' in the neighbourhood of the Ising critical point was efficiently reached using a parametric representation of the model in three dimensions. On the other hand, a similar approach is not the best choice in two dimensions: the degree of approximation inherited from the polynomial expansion of $h(\theta)$ \cite{Caselle:2000nn} results in numerical values of the universal amplitude ratios too far from the exact ones \cite{Delfino:1997ck}. The situation is solved by performing perturbative expansions around the exactly integrable directions $t=0$ and $H=0$ in the phase diagram \cite{Fonseca:2001dc}, which correspond to the singular limits $\eta=0, \pm \infty$. We refer to \cite{Caselle:2020tjz} for an exhaustive discussion about the derivation of the universal function $\Omega$ also in the bidimensional Ising model.

\providecommand{\href}[2]{#2}\begingroup\raggedright\endgroup


\begin{thebibliography}{10}

\bibitem{Caselle:2020tjz}
M.~Caselle and M.~Sorba, \emph{{Charting the scaling region of the Ising
  universality class in two and three dimensions}},
  \href{https://doi.org/10.1103/PhysRevD.102.014505}{\emph{Phys. Rev. D}
  {\bfseries 102} (2020) 014505}
  [\href{https://arxiv.org/abs/2003.12332}{{\ttfamily 2003.12332}}].

\bibitem{Kos:2016ysd}
F.~Kos, D.~Poland, D.~Simmons-Duffin and A.~Vichi, \emph{{Precision Islands in
  the Ising and $O(N)$ Models}},
  \href{https://doi.org/10.1007/JHEP08(2016)036}{\emph{JHEP} {\bfseries 08}
  (2016) 036} [\href{https://arxiv.org/abs/1603.04436}{{\ttfamily
  1603.04436}}].

\bibitem{Caselle:1997hs}
M.~Caselle and M.~Hasenbusch, \emph{{Universal amplitude ratios in the 3-D
  Ising model}}, \href{https://doi.org/10.1088/0305-4470/30/14/010}{\emph{J.
  Phys. A} {\bfseries 30} (1997) 4963}
  [\href{https://arxiv.org/abs/hep-lat/9701007}{{\ttfamily hep-lat/9701007}}].

\bibitem{Hasenbusch:2010ua}
M.~Hasenbusch, \emph{{Universal amplitude ratios in the three-dimensional Ising
  universality class}},
  \href{https://doi.org/10.1103/PhysRevB.82.174434}{\emph{Nucl. Phys. B}
  {\bfseries 82} (2010) 174434}
  [\href{https://arxiv.org/abs/1004.4983}{{\ttfamily 1004.4983}}].

\bibitem{Schofield:1969zza}
P.~Schofield, \emph{{Parametric Representation of the Equation of State Near A
  Critical Point}},
  \href{https://doi.org/10.1103/PhysRevLett.22.606}{\emph{Phys. Rev. Lett.}
  {\bfseries 22} (1969) 606}.

\bibitem{Campostrini:2002cf}
M.~Campostrini, A.~Pelissetto, P.~Rossi and E.~Vicari, \emph{{25th order high
  temperature expansion results for three-dimensional Ising like systems on the
  simple cubic lattice}},
  \href{https://doi.org/10.1103/PhysRevE.65.066127}{\emph{Phys. Rev. E}
  {\bfseries 65} (2002) 066127}
  [\href{https://arxiv.org/abs/cond-mat/0201180}{{\ttfamily
  cond-mat/0201180}}].

\bibitem{Tarko:1975zz}
H.B.~Tarko and M.E.~Fisher, \emph{{Theory of critical point scattering and
  correlations. III. The Ising model below $T_c$ and in a field}},
  \href{https://doi.org/10.1103/PhysRevB.11.1217}{\emph{Phys. Rev. B}
  {\bfseries 11} (1975) 1217}.

\bibitem{Lacey:2006bc}
R.A.~Lacey, N.N.~Ajitanand, J.M.~Alexander, P.~Chung, W.G.~Holzmann, M.~Issah
  et~al., \emph{{Has the QCD Critical Point been Signaled by Observations at
  RHIC?}}, \href{https://doi.org/10.1103/PhysRevLett.98.092301}{\emph{Phys.
  Rev. Lett.} {\bfseries 98} (2007) 092301}
  [\href{https://arxiv.org/abs/nucl-ex/0609025}{{\ttfamily nucl-ex/0609025}}].

\bibitem{Stephanov:2006zvm}
M.A.~Stephanov, \emph{{QCD phase diagram: An Overview}},
  \href{https://doi.org/10.22323/1.032.0024}{\emph{PoS} {\bfseries LAT2006}
  (2006) 024} [\href{https://arxiv.org/abs/hep-lat/0701002}{{\ttfamily
  hep-lat/0701002}}].

\bibitem{parotto2020qcd}
P.~Parotto, M.~Bluhm, D.~Mroczek, M.~Nahrgang, J.~Noronha-Hostler, K.~Rajagopal
  et~al., \emph{{QCD equation of state matched to lattice data and exhibiting a
  critical point singularity}},
  \href{https://doi.org/10.1103/PhysRevC.101.034901}{\emph{Phys. Rev. C}
  {\bfseries 101} (2020) 034901}
  [\href{https://arxiv.org/abs/1805.05249}{{\ttfamily 1805.05249}}].

\bibitem{Kadam:2020utt}
G.~Kadam, H.~Mishra and M.~Panero, \emph{{Critical exponents and transport
  properties near the QCD critical endpoint from the statistical bootstrap
  model}}, \href{https://doi.org/10.1140/epjc/s10052-021-09596-6}{\emph{Eur.
  Phys. J. C} {\bfseries 81} (2021) 795}
  [\href{https://arxiv.org/abs/2011.02171}{{\ttfamily 2011.02171}}].

\bibitem{Pan:2021svt}
X.~Pan, M.~Xu and Y.~Wu, \emph{{Cumulants and factorial cumulants in the
  three-dimensional Ising universality class}},
  \href{https://doi.org/10.1142/S0218301321500361}{\emph{Int. J. Mod. Phys. E}
  {\bfseries 30} (2021) 2150036}
  [\href{https://arxiv.org/abs/2101.02822}{{\ttfamily 2101.02822}}].

\bibitem{Pan:2021fus}
X.~Pan, \emph{{Fixed point behavior of cumulants in the three-dimensional Ising
  universality class}},  \href{https://arxiv.org/abs/2107.12758}{{\ttfamily
  2107.12758}}.

\bibitem{Kotov:2021hri}
A.Y.~Kotov, M.P.~Lombardo and A.~Trunin, \emph{{Gliding Down the QCD Transition
  Line, from $N_f$ = 2 till the Onset of Conformality}},
  \href{https://doi.org/10.3390/sym13101833}{\emph{Symmetry} {\bfseries 13}
  (2021) 1833} [\href{https://arxiv.org/abs/2111.00569}{{\ttfamily
  2111.00569}}].

\bibitem{Kotov:2021rah}
A.Y.~Kotov, M.P.~Lombardo and A.~Trunin, \emph{{QCD transition at the physical
  point, and its scaling window from twisted mass Wilson fermions}},
  \href{https://doi.org/10.1016/j.physletb.2021.136749}{\emph{Phys. Lett. B}
  {\bfseries 823} (2021) 136749}
  [\href{https://arxiv.org/abs/2105.09842}{{\ttfamily 2105.09842}}].

\bibitem{Caselle:2000nn}
M.~Caselle, M.~Hasenbusch, A.~Pelissetto and E.~Vicari, \emph{{The critical
  equation of state of the 2-D Ising model}},
  \href{https://doi.org/10.1088/0305-4470/34/14/302}{\emph{J. Phys. A}
  {\bfseries 34} (2001) 2923}
  [\href{https://arxiv.org/abs/cond-mat/0011305}{{\ttfamily
  cond-mat/0011305}}].

\bibitem{Delfino:1997ck}
G.~Delfino, \emph{{Universal amplitude ratios in the two-dimensional Ising
  model}}, \href{https://doi.org/10.1016/S0370-2693(97)01457-3}{\emph{Phys.
  Lett. B} {\bfseries 419} (1998) 291}
  [\href{https://arxiv.org/abs/hep-th/9710019}{{\ttfamily hep-th/9710019}}].

\bibitem{Fonseca:2001dc}
P.~Fonseca and A.~Zamolodchikov, \emph{{Ising field theory in a magnetic field:
  Analytic properties of the free energy}},
  \href{https://doi.org/10.1023/A:1022147532606}{\emph{J. Stat. Phys.}
  {\bfseries 110} (2003) 527}
  [\href{https://arxiv.org/abs/hep-th/0112167}{{\ttfamily hep-th/0112167}}].

\end{thebibliography}
\end{document}